\documentclass{ptephy_v1}
\usepackage{graphicx}      
\usepackage{pstricks}                                                                   
\usepackage{pstricks-add}
\newcommand{\Fh}[2]{\,{}_#1F_#2}
\newcommand{\Fs}[3]
{\!\!\left[\begin{array}{c}#1\,;\\#2\,;
\end{array}#3\right]}
\newcommand{\Fz}[3]{\Fs{#1}{#2}{#3}}
\begin{document}
\title{Hypergeometric presentation for one-loop 
contributing to $H\rightarrow Z\gamma$}
\author{Khiem Hong Phan and 
Dzung Tri Tran}
\affil{$^{1)}$University of Science Ho Chi Minh City,
227 Nguyen Van Cu, District 5, HCM City, Vietnam\\
$^{2)}$Vietnam National University Ho Chi Minh City, 
Linh Trung Ward, Thu Duc District, HCM City, Vietnam\\
\email{phkhiem@hcmus.edu.vn}}
\begin{abstract}%
In this paper, new analytic formulas for 
one-loop contributing to Higgs decay 
channel $H \rightarrow  Z\gamma$ are 
presented in terms of hypergeometric 
functions. The calculations are performed 
by following the technique for tensor 
one-loop reduction developed in 
[A.~I.~Davydychev, Phys.\ Lett.\ B {\bf 263} 
(1991) 107]. For the first time, one-loop 
form factors for the decay process are 
shown which are valid at arbitrary 
space-time dimension $d$. 
\end{abstract}
\subjectindex{B87}
\maketitle
\section{Introduction}
Among Higgs ($H$) decay processes, the decay 
channel $H \rightarrow Z\gamma$ is the most 
important at the Large Hadron Collider 
(LHC)~\cite{Abazov:2008wg,Chatrchyan:2013vaa,
Aaboud:2017uhw}. Because the channel arises 
at first from one-loop Feynman diagrams. 
As a result, the decay width of this channel is
sensitive to new physics in which we assume 
that new heavy particles may exchange 
in one-loop diagrams. For this reason, theoretical 
evaluations for one-loop and higher-loop decay 
amplitudes of $H\rightarrow Z\gamma$ play crucial 
roles in controlling standard model (SM) 
background as well as constraining physical 
parameters in many beyond standard models (BSM).

There have been many computations for one-loop 
contributions to $H \rightarrow Z \gamma$ 
within SM and its extensions
in~\cite{Cahn:1978nz,Bergstrom:1985hp,Martinez:1989kr,
Djouadi:1996yq,Djouadi:1997yw,Chiang:2012qz,Chen:2013vi,
Cao:2013ur,Bonciani:2015eua, Hammad:2015eca,Belanger:2014roa,  
No:2016ezr, Monfared:2016vwr, Fontes:2014xva,
Funatsu:2015xba, Hue:2017cph, Hung:2019jue,
Boradjiev:2017khm}. The calculations have performed 
following the method for tensor one-loop reduction 
in~\cite{Denner:1991kt}. When one-loop 
contributions to $H\rightarrow Z \gamma$ are 
evaluated in unitary gauge,
the results may meet large numerical 
cancellations. This is because higher-rank tensor 
one-loop integrals appears from Feynman loop 
diagrams with exchanging by vector bosons. To 
avoid this problem, many of the above references 
have considered the calculations in 't Hooft-Feynman 
gauge. In this gauge, we need to handle more 
Feynman diagrams with exchanging by Goldstone 
bosons. As a result, the calculations  are rather  
complicated. Furthermore, when we consider two-loop 
or higher-loop corrections to $H\rightarrow Z \gamma$, 
two-loop and higher-loop Feynman integrals may be 
evaluated by applying 
methods~\cite{IBP,Laporta:2001dd,Tarasov:1996br},
the resulting integrals may contain the one-loop 
integrals in general space-time dimension. These 
integrals have been not available in previous papers.

In this paper, we apply
an alternative approach for evaluating one-loop 
contributions to $H\rightarrow Z \gamma$. 
In this calculation, we follow the method for 
tensor one-loop reduction developed 
in~\cite{Davydychev:1991va} in which 
tensor integrals are decomposed into 
scalar functions with arbitrary propagator 
indexes and at higher space-time dimension 
$d>4$. Using integration-by-part method 
(IBP)~\cite{IBP, Laporta:2001dd}, 
scalar one-loop integrals are then expressed in 
terms of master integrals which can be solved 
analytically via generalized hypergeometric 
series. For instant, analytic formulas for 
the master integrals which are one-loop one-, 
two-, three-point functions at general 
$d$ appearing in $H\rightarrow Z \gamma$ 
are provided in this work. Therefore,
our methods are easy to 
apply for $H\rightarrow Z \gamma$ and 
expect to be numerical stability in unitary gauge. 
Furthermore, our analytic expressions for 
the form factors of the decay process are 
general as well as valid at 
arbitrary space-time dimension.

The layout of the paper is as follows: In section 2,
we present a general method for evaluating one-loop
Feynman integrals. Using the method, the computations
for one-loop contributions to Higgs decay to $Z$ 
photon are reported in the section $3$. Conclusions 
are shown in section $4$. Several useful formulas used 
in this calculation and detailed calculations 
for the process amplitudes are given in the 
appendixes.
\section{Method \label{method}}   
In this section, we describe a general 
approach for evaluating one-loop Feynman integrals.
In general, tensor one-loop $N$-point Feynman 
integrals with rank $M$ are defined as follows: 
\begin{eqnarray}
&& \hspace{-1cm}
J_{N,\mu_1\mu_2\cdots\mu_M} 
(d;\{\nu_1,\nu_2,\cdots, \nu_N\}) 
\equiv 
J_{N,\mu_1\mu_2\cdots\mu_M} 
(d;\{\nu_1,\nu_2,\cdots, \nu_N\}; \{p_ip_j; m_i^2\}) 
= \\
&=& \int \frac{d^d k}{i\pi^{d/2}} 
\dfrac{k_{\mu_1}k_{\mu_2}\cdots k_{\mu_M}}
{[(k+q_1)^2 -m_1^2 + i\rho]^{\nu_1}
[(k+q_2)^2 -m_2^2 + i\rho]^{\nu_2}
\cdots 
[(k+q_N)^2 -m_N^2 + i\rho]^{\nu_N}}. 
\nonumber
\end{eqnarray}
Where $p_i$ ($m_i$) for $i=1,2,\cdots, N$ 
are external momenta (internal masses) 
respectively. In this convention, 
$q_1 =p_1, q_2 =p_1+p_2, \cdots, 
q_i = \sum_{j=1}^{i}p_j,$ and 
$q_N=\sum_{j=1}^{N}p_j=0$ thanks to momentum 
conservation. The term $i\rho$ is Feynman's 
prescription and $d$ is space-time dimension.  
One of physical interests is 
$d= 4+2n-2\epsilon$ for $n\in \mathbb{N}$.

Following the method for tensor reduction 
in Ref.~\cite{Davydychev:1991va}, tensor 
one-loop integrals can be reduced to scalar 
functions with the shifted space-time 
dimension as follows: 
\begin{eqnarray}
&& \hspace{-0.7cm}
J_{N,\mu_1\mu_2\cdots\mu_M} 
(d;\{\nu_1,\nu_2,\cdots, \nu_N\}) 
=
\sum\limits_{\lambda, \kappa_1,\kappa_2 
\cdots, \kappa_N} 
\left(-\frac{1}{2}\right)^{\lambda} 
\Big\{[g]^\lambda [q_1]^{\kappa_1} 
[q_2]^{\kappa_2}\cdots [q_N]^{\kappa_N} \Big\}_
{\mu_1\mu_2\cdots\mu_M}    \nonumber\\
&&\hspace{1cm}
\times (\nu_1)_{\kappa_1}(\nu_2)_{\kappa_2}
\cdots (\nu_N)_{\kappa_N}
\;\;
J_N (d+2(M-\lambda);
\{\nu_1+\kappa_1,\nu_2+\kappa_2,
\cdots, \nu_N+\kappa_N\}).
\nonumber\\
\end{eqnarray}
Here $\lambda, \kappa_1, \kappa_2, \cdots, \kappa_N$ 
satisfy the following constrains 
$2\lambda +\kappa_1+\kappa_2+\cdots+\kappa_N=M$,
$0 \leq \kappa_1, \kappa_2,\cdots, \kappa_N \leq M$ 
and $0 \leq \lambda \leq [M/2]$ 
(integer of $M/2$). The Pochhammer symbol is used
as $(a)_\kappa = \Gamma(a+\kappa)/\Gamma(a)$. 
The tensor $\{[g]^\lambda [q_1]^{\kappa_1} [q_2]^{\kappa_2} 
\cdots [q_N]^{\kappa_N} \}_{\mu_1\mu_2\cdots \mu_M} $ 
is symmetric in regard to
$\mu_1, \mu_2,\cdots, \mu_M$. 
It is formed from $\lambda$ of metric $g_{\mu\nu}$, 
$\kappa_1$ of momentum $q_1$, $\cdots$,  
$\kappa_N$ of momentum $q_N$.
The $J_N(d+2(M-\lambda);\{\nu_1+\kappa_1,
\nu_2+\kappa_2,\cdots, \nu_N+\kappa_N\})$ 
with changing space-time dimension to 
$d+ 2(M-\lambda)$, raising powers of 
propagators $\{\nu_i+\kappa_i\}$ for $i=1,2,\cdots, N$
are scalar one-loop $N$-point functions.

In the next step, the scalar integrals 
$J_N(d; \{\nu_1,\nu_2,\cdots,\nu_N\} )$ 
are casted into subset of master functions 
by using IBP~\cite{IBP}. In detail, 
applying the operator 
$\frac{\partial}{\partial k} \cdot k$ 
to the integrand of $J_N(d; 
\{\nu_1,\nu_2,\cdots,\nu_N\} )$
and setting $k$ to be the 
momentum of $N$ internal lines ($k\equiv\{k+q_1, 
k+q_2,\cdots, k+q_N\}$). As a result, 
$J_N(d; \{\nu_1,\nu_2,\cdots,\nu_N\} )$
can be expressed in terms of
$J_N(d;\{1,1,\cdots, 1\})$ and 
$J_{N-1}(d; \{\nu'_1,\nu'_2,\cdots,\nu'_{N-1}\} )$.
In this recurrence way \cite{Laporta:2001dd},
we arrive at the master integrals
which can be solved analytically. For examples, 
they may be $J_N(d;\{1,1,\cdots, 1\})$ 
and $J_{N-L}(d; \{\nu''_1,\nu''_2,\cdots,
\nu''_{N-L}\} )$ with $L<N$. Recently, scalar
one-loop integrals at general $d$ have been 
expressed in terms 
of generalized hypergeometric 
series~\cite{Bluemlein:2015sia,Phan:2019qee,
Phan:2018cnz,Bluemlein:2017rbi}.

In the Appendix $B$, this method 
is demonstrated in detail for the case of 
$H\rightarrow  Z\gamma$. We show here all 
analytic results for the master integrals 
involving the decay process. In particular,
scalar one-loop one-point functions
with arbitrary propagator index $\nu$ 
are given~\cite{tHooft:1978jhc}:
\begin{eqnarray} 
J_1 (d;\{ \nu \}; M^2)
&=& 
(-1)^{\nu}
\dfrac{\Gamma(\nu - d/2)}
{\Gamma(\nu)}\;(M^2)^{d/2 - \nu}.
\end{eqnarray}
Scalar one-loop two-point functions
with general propagator indexes 
$\nu_1, \nu_2$ in the case of 
$m_1^2=m_2^2=M^2$ read~\cite{khiemcadj}:
\begin{eqnarray}
J_2(d;\{\nu_1,\nu_2 \}; p^2, M^2) 
&=& 
(-1)^{N_2}
\dfrac{\Gamma(N_2-d/2)}{\Gamma(N_2)}
(M^2)^{d/2-N_2}\;
\Fh32\Fz{\nu_1, \nu_2,N_2-d/2 }
{\frac{N_2}{2}, 
\frac{N_2+1}{2} }
{\dfrac{p^2}{4M^2} }.
\nonumber\\
\label{3F2-J2}
\end{eqnarray}
Here $N_2 = \nu_1+\nu_2$, 
$p^2 =0, M_H^2, M_Z^2$ and 
$M^2 =m_f^2, M_W^2$ in 
this calculation. Other master
integrals which are scalar one-loop
three-point functions are given:
\begin{eqnarray}
\dfrac{
J_3(d;\{ 1, 1, 1 \}; p_2^2, M_H^2, M^2) 
}{\Gamma \left( 2 - d/2 \right)  }
&=&\dfrac{ (d - 4) M_H^2 }{
4 (M_H^2  - p_2^2)} (M ^2)^{d/2 - 3}  \times 
\\
&&\hspace{-2cm} \times
\left\{  
\Fh32\Fz{1, 1, 3 - d/2 }{3/2, 2}{
\dfrac{M_H^2}{4 M^2} }  
-
\Fh32\Fz{1, 1, 3 - d/2 }{3/2, 2}{
\dfrac{p_2^2}{4 M^2} } 
\right\}, \nonumber
\end{eqnarray}
\begin{eqnarray}
\dfrac{J_3(d;\{ 1, 2, 1 \}; p_2^2, M_H^2, M^2) 
}{\Gamma \left( 2 - d/2 \right)  }
&=&  \dfrac{\left( 4 - d \right)}{
2 (M_H^2-p_2^2)}
(M ^2)^{d/2 - 3} 
\times \\
&&\hspace{-2cm} \times
\left\{ 
\Fh32\Fz{1, 2, 3 - d/2 }{3/2, 2}{
\dfrac{M_H^2}{4 M^2} }
-
\Fh32\Fz{1, 2, 3 - d/2 }{3/2, 2}{
\dfrac{p_2^2}{4 M^2} }  
\right\}, 
\nonumber
\end{eqnarray}
\begin{eqnarray}
\dfrac{
J_3(d;\{ 1, 3, 1 \}; p_2^2, M_H^2, M^2) 
}{\Gamma \left( 2 - d/2 \right)  }
&=& 
\dfrac{\left( 6 - d \right) 
\left( d - 4 \right)
}{16(M_H^2 - p_2^2)} (M ^2)^{d/2 - 4}  
\times 
\\
&&
\hspace{-2cm} \times 
\left\{ 
\Fh32\Fz{1, 2, 4 - d/2 }{3/2, 2}{
\dfrac{M_H^2}{4 M^2} }
-
\Fh32\Fz{1, 2, 4 - d/2 }{3/2, 2}{
\dfrac{p_2^2}{4 M^2} }  
\right\}, 
\nonumber
\end{eqnarray}
\begin{eqnarray}
\hspace{-3cm}
\dfrac{
J_3(d;\{ 2, 2, 1 \}; p_2^2, M_H^2, M^2) 
}{\Gamma \left( 2 - d/2 \right)  }
&=&
\left( 4 - d \right) 
(M^2) ^{d/2 - 4} \times \\
&& 
\hspace{-5.7cm} \times
\Bigg\{  \hspace{0.1cm}
\dfrac{\left( 6 - d \right)  
M_H^2 \left(4 M^2  - M_H^2 \right)}
{ 16 M^2 (M_H^2-p_2^2)^2 }
\Fh32\Fz{1, 2, 4 - d/2 }{3/2, 2}
{\dfrac{M_H^2}{4 M^2} } 
\nonumber\\
&& 
\hspace{-5.3cm}
\left. + \dfrac{ \left( 6 - d \right) 
\left[ M_H^2 p_2^2 - 
2 M^2 (M_H^2 + p_2^2) \right] }
{ 16 M^2 (M_H^2-p_2^2)^2}
\Fh32\Fz{1, 2, 4 - d/2 } {3/2, 2}
{\dfrac{p_2^2}{4 M^2} } \right. 
\nonumber 
\end{eqnarray}
\begin{eqnarray}
&&
-\dfrac{ \left(6 - d \right) M_H^2 }
{ 4 (M_H^2-p_2^2)^2 } 
\Fh32\Fz{1, 2, 3 - d/2 }{3/2 , 2}
{ \dfrac{M_H^2}{4 M^2} }
\nonumber\\
&&
-\dfrac{ \left[ 2 M_H^2 (d - 5) - 2 p_2^2 \right]}
{ 8 (M_H^2-p_2^2)^2 }
\Fh32\Fz{ 1, 2, 3 - d/2 }{ 3/2 , 2 }
{ \dfrac{p_2^2}{4 M^2} }  
\nonumber\\
&&
+   
\dfrac{ \left(d - 4 \right) M_H^2}
{ 8 (M_H^2-p_2^2)^2}
\Fh32\Fz{1, 1, 3 - d/2}{ 
3/2, 2}{ \dfrac{M_H^2}{4 M^2} } 
+ \dfrac{ \left(4 - d \right) 
p_2^2}{
8 (M_H^2-p_2^2)^2}
\Fh32\Fz{1, 1, 3 - d/2}{
3/2, 2}
{ \dfrac{p_2^2}{4 M^2} } 
\Bigg\}. 
\nonumber
\end{eqnarray}
Where $p_2^2 = M_Z^2, 0$ and $M^2 =m_f^2, M_W^2$ 
in the present calculation.

We are going to apply this method for evaluating
the Higgs decay processes. The first results for
one-loop
contributions to $H\rightarrow \gamma\gamma$
have been published in \cite{khiemcadj}. 
In the next section, we show new analytic 
results for $H\rightarrow Z\gamma$ by mean of 
$_3F_2$ hypergeometric series. 
\section{Hypergeometric presentation   
for one-loop contributing to           
$H\rightarrow Z\gamma$}                
\begin{figure}[ht]
\begin{center}
\includegraphics
[width=6in,height=7cm,angle=0]
{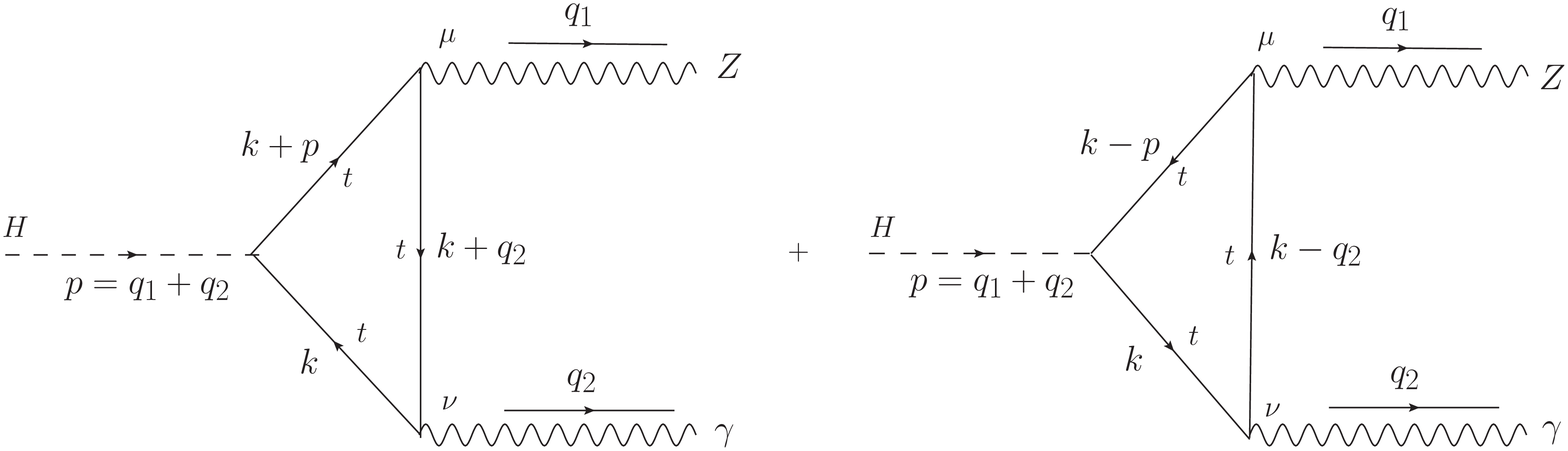}
\end{center}
\caption{Feynman diagrams contributing 
to the $H \longrightarrow Z \gamma$ decay 
through top quark loop in unitary gauge.}
\label{top_diagrams}
\end{figure}

\begin{figure}[ht]
\begin{center}
\includegraphics[width=6in,height=7cm,angle=0]
{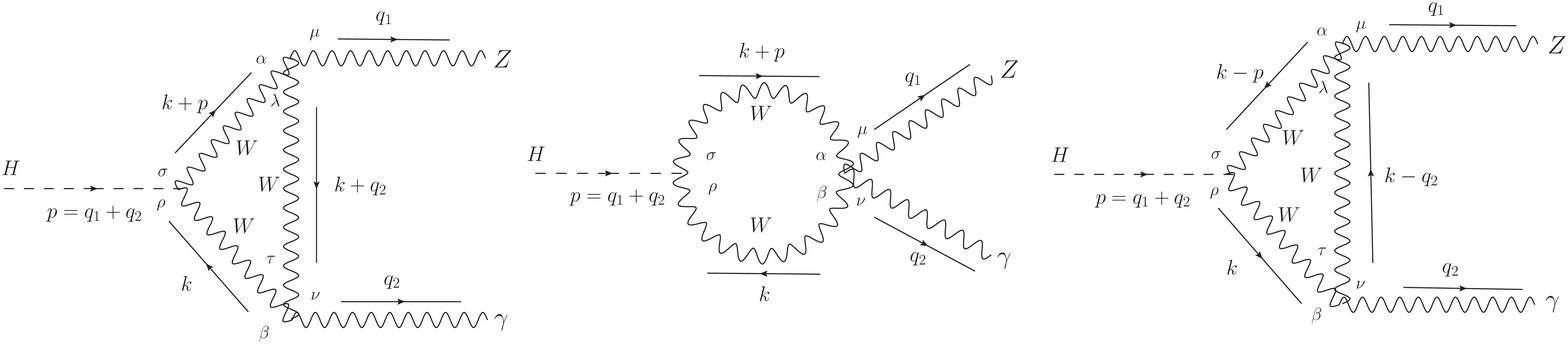}
\end{center}
\caption{Feynman diagrams contributing 
to the $H \rightarrow Z \gamma$ decay 
through W boson loop in unitary gauge.
\label{wboson_diagrams}}
\end{figure}

In unitary gauge, the decay process 
$H\rightarrow Z\gamma$ consists top loop
and $W$ boson loop as shown in 
Figs.~{\ref{top_diagrams}, \ref{wboson_diagrams}}.
In general, the total amplitude of the 
decay $H \rightarrow Z \gamma$ is expressed 
in terms of form factors with reflecting 
the Lorentz invariant structure and 
the content of gauge symmetry as follows:
\begin{eqnarray}
\label{total amplitude}
i \mathcal{A} _{H \rightarrow Z \gamma}
&=&
i \mathcal{A} _{\mu \nu} \, 
\varepsilon ^{\mu *} _1 (q_1)  
\varepsilon ^{\nu *} _2 (q_2) =
\\
&=&\Big( 
F_{00} \, g_{\mu \nu} 
+ \sum \limits _{i,j = 1} ^2 
F_{i j} \, q_{i,\mu} q_{j,\nu} 
+ F_5 \times 
i \epsilon _{\mu \nu \alpha \beta} 
q_1^\alpha q_2^\beta 
\Big)\, 
\varepsilon ^{\mu *} _1 (q_1)  
\varepsilon ^{\nu *} _2 (q_2).  
\nonumber
\end{eqnarray}
Where $\varepsilon ^{\mu *} _1$ 
and $\varepsilon ^{\nu *} _2$ are the 
polarization vectors of the $Z$ 
boson and the photon $\gamma$ 
respectively. 
$\epsilon _{\mu \nu \alpha \beta}$ 
is the Levi-Civita  tensor. 
Kinematic invariant variables
related to this process are
\begin{eqnarray}
q_1^2 = M_Z^2, \quad q_2 ^2 = 0,\quad 
p^2 = (q_1 + q_2)^2 = M_H^2.
\end{eqnarray}
We also have 
$\varepsilon ^{\nu *} _2 (q_2) q_{2,\nu} = 0$
for external photon. Following Ward identity, 
we confirm that 
\begin{eqnarray}
F_{11} = 0, \quad  F_{00}
= - \left( q_1 \cdot q_2 \right) F_{21}
= \frac{M_Z^2 - M_H^2}{2}\;F_{21}
\end{eqnarray}
and $F_{12,22}$ do not contribute 
to the total amplitude. Summing all the 
top-loop diagrams, the result shows 
that $F_{5} = 0$. Detailed calculations 
for the form factors at general $d$ are 
presented in the appendix $D$.
The total amplitude for this decay 
process is then casted in the form 
of
\begin{eqnarray}
i \mathcal{A} _{H \rightarrow Z \gamma}
= \dfrac{e^3}{ \sin \theta_W M_W} 
\mathcal{F} _{H \rightarrow Z \gamma} 
(d; M_H^2, M_Z^2, M_W^2, m_f^2) 
\left[ q_{2,\mu} q_{1,\nu} - 
\left( q_1 \cdot q_2 \right) 
\, g_{\mu \nu} \right] 
\varepsilon ^{\mu *} _1 (q_1) 
\varepsilon ^{\nu *} _2 (q_2),
\nonumber\\
\end{eqnarray}
where $\mathcal{F} _{H \rightarrow Z \gamma} 
(d; M_H^2, M_Z^2, M_W^2, m_f^2)$ 
are form factors which can be derived from 
$F_{00}$ or $F_{21}$. These form factors are 
decomposed in terms of $W$-loop and 
top-loop (including fermion-loop) 
contributions as follows:
\begin{eqnarray}
\mathcal{F} _{H \rightarrow Z \gamma} 
(d; M_H^2, M_Z^2, M_W^2, m_f^2)
&=&
\cot \theta_W \, 
\mathcal{F} ^{(W)} _{H \rightarrow Z \gamma} 
(d; M_H^2, M_Z^2, M_W^2) 
\\
&&
+ \sum \limits _f 
\dfrac{Q_f N_C }{e} 
\left( \lambda^f_1 
+ \lambda^f_2 \right) 
\mathcal{F} ^{(f)} _{H \rightarrow Z \gamma} 
(d; M_H^2, M_Z^2, m_f^2).
\nonumber
\end{eqnarray}
Where $\theta _W$ is Weinberg angle, 
$I^{3}_f$, $Q_f$ and $m_f$ are iso-spin, 
electric charge, mass of fermions 
$f$ in the loops respectively.
$N_C$ is a color factor for fermions. 
It becomes $1$ for leptons and $3$ for 
quarks. We use the symbolic-manipulation 
{\tt Package-X}~\cite{Patel:2015tea} 
to handle all
Dirac and tensor algebra in $d$ dimension.
\subsection{Form factors}           
We show two representations for the
form factors in terms of $_3F_2$ 
hypergeometric functions in this 
subsection.
\subsubsection{First representation} 
We first present the form factors 
which are derived from $F_{00}$ in 
(\ref{total amplitude}) in terms of 
$_3F_2$ hypergeometric functions as 
follows:  
\begin{eqnarray}
\label{Fw1}
&&\hspace{-0.8cm}
\dfrac{
\mathcal{F}_{H \rightarrow Z \gamma} ^{(W)} 
(d; M_H^2, M_Z^2,M_W^2)} 
{\Gamma \left( 2 - d/2 \right)}
= 
\dfrac{  (M_W^2)^{d/2 - 2}  }
{ (4 \pi)^{d/2} \; M_W^2
(M_Z^2  - M_H^2)^2 }
\; \times 
\\
&&
\times
\Bigg\{
\hspace{0.1cm}
\left( 4 - d \right) 
(M_Z^2 - 4 M_W^2 ) ( M_H^2 -M_Z^2 ) 
\times \nonumber \\
&&
\hspace{0.7cm}
\times \left(  M_H^2 
\Fh32\Fz{1, 1, 3 - d/2 }
{3/2, 2}
{\dfrac{M_H^2}{4 M_W^2} }
- M_Z^2 \Fh32\Fz{1, 1, 3 - d/2 }
{3/2, 2}
{\dfrac{M_Z^2}{4 M_W^2} } \right)
\nonumber \\
&&
\hspace{0.5cm}
\left. + \Big[ 2 M_W^2 ( M_H^2 - M_Z^2 ) 
-  M_H^2  M_Z^2 + 4 M_W^4 ( d - 1)  \Big] 
\right. \times  
\nonumber \\
&& 
\hspace{0.7cm} 
\times \Bigg(\; 
(M_H^2  -M_Z^2) 
\Fh32\Fz{ 2, 1, 2 - d/2}
{3/2 , 2 }
{\dfrac{M_H^2}{4 M_W^2} }
\nonumber
\\
&& 
\hspace{1.5cm}
+ M_Z^2 \Fh32\Fz{1, 1, 2 - d/2}
{ 3/2, 2}
{\dfrac{M_Z^2}{4 M_W^2} }
- M_H^2 \Fh32\Fz{1, 1, 2 - d/2}
{3/2, 2}
{\dfrac{M_H^2}{4 M_W^2} }   \Bigg) 
\hspace{0.1cm}
\Bigg\} ,
\nonumber
\end{eqnarray}
and
\begin{eqnarray}
\label{Ft1}
&& 
\dfrac{ 
\mathcal{F}_{H \rightarrow Z \gamma} ^{(t)}  
(d; M_H^2, M_Z^2,m_t^2)}
{\Gamma 
\left( 2 - d/2 \right)}
=
\dfrac{(m_t^2)^{d/2 - 2} }{ (4 \pi)^{d/2} \;
(M_Z^2 - M_H^2)^2  }
\times
\\
&&\times \Bigg\{
\hspace{0.2cm}
 \left(4-d \right) M_H^2 (M_H^2 - M_Z^2) 
\Fh32\Fz{1, 1, 3 - d/2}{3/2, 2}
{\dfrac{M_H^2}{4 m_t^2}} + 8 M_H^2 m_t^2 
\Fh32\Fz{1, 1, 2 - d/2}
{3/2, 2}{ \dfrac{M_H^2}{4 m_t^2} }
\nonumber\\
&&
\hspace{0.7cm} \left. 
+ \left(4-d \right) M_Z^2 (M_Z^2 - M_H^2) 
\Fh32\Fz{1, 1, 3 - d/2}
{ 3/2, 2}
{ \dfrac{M_Z^2}{4 m_t^2} } 
- 8 M_Z^2 m_t^2 
\Fh32\Fz{1, 1, 2 - d/2 }
{ 3/2, 2}
{ \dfrac{M_Z^2}{4 m_t^2} } 
\right. \nonumber \\
&&\hspace{0.7cm} 
 - 8 (M_H^2 - M_Z^2) m_t^2
\Fh32\Fz{ 1, 1, 2 - d/2 }
{ 3/2 , 1 }
{ \dfrac{M_H^2}{4 m_t^2} }
\hspace{0.1cm}
\Bigg \}.
\nonumber
\end{eqnarray}
The form factors $\mathcal{F}_{H \rightarrow 
Z \gamma}^{(f)} (d; M_H^2, M_Z^2,m_f^2)$ 
are obtained by 
replacing $m_t \rightarrow m_f$ 
in Eq. (\ref{Ft1}).
For the form factors which have 
fermion masses are smaller than $M_H/2$, 
the argument of hypergeometric 
functions $_3F_2$ is greater than $1$
(or $|M_H^2/4m_f^2|>1$).
We subsequently apply analytic 
continuation in Eq. (\ref{analytic3F2}) 
for $_3F_2$ appearing 
in the form factors $\mathcal{F}_{H \rightarrow 
Z \gamma}^{(f)} (d; M_H^2, M_Z^2,m_f^2)$.
In the limit $d \rightarrow 4$, 
we confirm that the terms 
in curly brackets of 
right hand side results 
of (\ref{Fw1},\ref{Ft1}) tend to zero
\begin{eqnarray}
\Big[ 2 M_W^2 ( M_H^2 - M_Z^2 ) 
-  M_H^2  M_Z^2 + 12 M_W^4  \Big] 
\Big( M_Z^2 + (M_H^2  -M_Z^2) - M_H^2 \Big) &=&0,
\\
8 M_H^2 m_t^2 
- 8 M_Z^2 m_t^2 
- 8 (M_H^2 - M_Z^2) m_t^2  &=& 0.
\end{eqnarray}
It means that the form factors 
always stay finite in the limit.
\subsubsection{Second representation}   
Another presentation for the form factors
which are obtained from $F_{21}$ in 
(\ref{total amplitude}) are given:
\begin{eqnarray}
\label{Fw2}
&&\dfrac{
\mathcal{F}_{H \rightarrow Z \gamma} ^{(W)} 
(d; M_H^2, M_Z^2,M_W^2)} 
{\Gamma \left( 2 - d/2 \right)}
= 
\dfrac{  (M_W^2)^{d/2 - 2}  }
{ (4 \pi)^{d/2} \; M_W^4
(M_Z^2  - M_H^2)^2 }
\;
\times 
\\
&&
\times
\Bigg\{
\hspace{0.1cm}
\left( 4 - d \right) 
M_W^2
(M_Z^2 - 4 M_W^2 ) ( M_H^2 -M_Z^2 ) 
\times \nonumber \\
&&
\hspace{0.5cm}
\times 
\left(  M_H^2 
\Fh32\Fz{1, 1, 3 - d/2 }
{3/2, 2}
{\dfrac{M_H^2}{4 M_W^2} }
- M_Z^2 
\Fh32\Fz{1, 1, 3 - d/2 }
{3/2, 2}
{\dfrac{M_Z^2}{4 M_W^2} } \right)
\nonumber \\
&&
\hspace{0.5cm}
\left. + \Big[ 2 M_W^4 ( M_H^2 - M_Z^2 ) 
-  M_H^2 M_W^2 M_Z^2 + 4 M_W^6 ( d - 1)  \Big] 
\right. \times  
\nonumber \\
&& 
\hspace{0.5cm} 
\times \Bigg[
\dfrac{  M_H^2 \left( 6 M_W^2 - M_H^2 \right) 
- 2 M_W^2 M_Z^2 }{2 M_W^2 } 
\Fh32\Fz{2, 1, 2 - d/2 }
{3/2, 2}
{\dfrac{M_H^2}{4 M_W^2} }
\nonumber\\
&& 
\hspace{0.8cm}
+ \dfrac{  M_H^2 ( M_Z^2 - 4 M_W^2) }{2 M_W^2 } 
\Fh32\Fz{2, 1, 2 - d/2 }
{3/2, 2}
{\dfrac{M_Z^2}{4 M_W^2} }
\nonumber\\
&& \hspace{0.8cm}
- 2 M_H^2 \left( 
\Fh32\Fz{2, 1, 2 - d/2 }
{3/2, 2}
{\dfrac{M_H^2}{4 M_W^2} } 
- \dfrac{M_H^2}{6 M_W^2} 
\Fh32\Fz{3, 2, 2 - d/2 }
{5/2, 3}
{\dfrac{M_H^2}{4 M_W^2} }
 \right)
\nonumber\\
&& \hspace{0.8cm}
+ \dfrac{2 M_H^2 (d - 1) 
- 2 M_Z^2}{ \left( d - 2 \right)} 
\left(  
\Fh32\Fz{2, 1, 2 - d/2 }
{3/2, 2}
{\dfrac{M_Z^2}{4 M_W^2} }
- \dfrac{M_Z^2}{6 M_W^2} 
\Fh32\Fz{3, 2, 2 - d/2 }
{5/2, 3}
{\dfrac{M_Z^2}{4 M_W^2} }
\right)
\nonumber\\
&& \hspace{0.8cm}
+ \dfrac{ d}{ \left( 2 - d \right)} 
\left( M_H^2 \, 
\Fh32\Fz{1, 1, 2 - d/2 }
{3/2, 2}
{\dfrac{M_H^2}{4 M_W^2} }
- \dfrac{M_H^4}{12 M_W^2} 
\Fh32\Fz{2, 2, 2 - d/2 }
{5/2, 3}
{\dfrac{M_H^2}{4 M_W^2} }
\right)
\nonumber\\
&& 
\hspace{0.8cm}
+ \dfrac{d}{\left( d - 2 \right)} \left( M_Z^2 \, 
\Fh32\Fz{1, 1, 2 - d/2 }
{3/2, 2}
{\dfrac{M_Z^2}{4 M_W^2} }
- \dfrac{M_Z^4}{12 M_W^2} 
\Fh32\Fz{2, 2, 2 - d/2 }
{5/2, 3}
{\dfrac{M_Z^2}{4 M_W^2} }
\right)
 \Bigg]
\hspace{0.2cm}
\Bigg\}
,
\nonumber
\end{eqnarray}
and
\begin{eqnarray}
\label{Ft2}
&& \hspace{-0.5cm} 
\dfrac{ 
\mathcal{F}_{H \rightarrow Z \gamma}^{(t)}  
(d; M_H^2, M_Z^2,m_t^2)}
{\Gamma 
\left( 2 - d/2 \right)}
=
\dfrac{(m_t^2)^{d/2 - 2} }{ (4 \pi)^{d/2} \;
(M_Z^2 - M_H^2)^2  }
\times
\\
&&\hspace{-0.4cm} 
\times 
\Bigg\{
\hspace{0.2cm}
\dfrac{4 M_H^2 \left(4 m_t^2  - M_H^2 \right)}{3} 
\Bigg(
\Fh32\Fz{2, 2, 2 - d/2}{5/2, 2}
{\dfrac{M_H^2}{4 m_t^2}}
+  2 \; 
\Fh32\Fz{3, 1, 2 - d/2}{5/2, 2}
{\dfrac{M_H^2}{4 m_t^2}}
\Bigg)
\nonumber\\
&&\hspace{-0.2cm} \left. 
+  \dfrac{4 M_H^2 M_Z^2 - 8 m_t^2 (M_H^2 + M_Z^2)}{3} 
\Bigg(
\Fh32\Fz{2, 2, 2 - d/2}
{ 5/2, 2}
{ \dfrac{M_Z^2}{4 m_t^2} } 
+  2 \; 
\Fh32\Fz{3, 1, 2 - d/2}
{ 5/2, 2}
{ \dfrac{M_Z^2}{4 m_t^2} } 
\Bigg)
\right. 
\nonumber \\
&&\hspace{-0.2cm} \left. 
- 16 M_H^2 m_t^2 \left( 
\Fh32\Fz{2, 1, 2 - d/2}{3/2, 2}
{\dfrac{M_H^2}{4 m_t^2}}
- \dfrac{M_H^2}{6 m_t^2} 
\Fh32\Fz{3, 2, 2 - d/2}{5/2, 3}
{\dfrac{M_H^2}{4 m_t^2}}
\right)
\right. 
\nonumber \\
&&\hspace{-0.2cm} \left. 
+  \dfrac{16 M_H^2 m_t^2 (d - 1)
-  16 M_Z^2 m_t^2}{ \left( d - 2 \right)} \left( 
\Fh32\Fz{2, 1, 2 - d/2}
{ 3/2, 2}
{ \dfrac{M_Z^2}{4 m_t^2} }
- \dfrac{M_Z^2}{6 m_t^2} 
\Fh32\Fz{3, 2, 2 - d/2}
{ 5/2, 3}
{ \dfrac{M_Z^2}{4 m_t^2} }
\right)
\right. 
\nonumber 
\end{eqnarray}
\begin{eqnarray}
&&\hspace{-0.2cm} \left. 
+ \dfrac{ 8 M_H^2 m_t^2 \, d}
{ \left( 2 - d \right)} \left( 
\Fh32\Fz{1, 1, 2 - d/2}{3/2, 2}
{\dfrac{M_H^2}{4 m_t^2}}
- \dfrac{M_H^2}{12 m_t^2} 
\Fh32\Fz{2, 2, 2 - d/2}{5/2, 3}
{\dfrac{M_H^2}{4 m_t^2}}
\right)  
\right. 
\nonumber \\
&&\hspace{-0.2cm} \left. 
+ \dfrac{ 8 M_Z^2 m_t^2 \, d}
{ \left( d - 2 \right)} \left( 
\Fh32\Fz{1, 1, 2 - d/2}
{ 3/2, 2}
{ \dfrac{M_Z^2}{4 m_t^2} }
- \dfrac{M_Z^2}{12 m_t^2} 
\Fh32\Fz{2, 2, 2 - d/2}
{ 5/2, 3}
{ \dfrac{M_Z^2}{4 m_t^2} }
\right)
\right. 
\nonumber \\
&&\hspace{-0.2cm} \left. 
+ 8 m_t^2 (M_H^2 - M_Z^2 ) \Bigg(
\Fh32\Fz{2, 1, 2 - d/2}{3/2, 2}
{\dfrac{M_H^2}{4 m_t^2}}
- 
\Fh32\Fz{2, 1, 2 - d/2}
{ 3/2, 2}
{ \dfrac{M_Z^2}{4 m_t^2} }
\Bigg)
\right. 
\nonumber \\
&&\hspace{-0.2cm} \left. 
+ \left( d - 4 \right) (M_H^2  -M_Z^2) \Bigg( M_H^2 \; 
\Fh32\Fz{1, 1, 3 - d/2}{3/2, 2}
{\dfrac{M_H^2}{4 m_t^2}}
- M_Z^2 \; 
\Fh32\Fz{1, 1, 3 - d/2}
{ 3/2, 2}
{ \dfrac{M_Z^2}{4 m_t^2} } 
\Bigg)
\right. 
\hspace{0.1cm}
\Bigg \}. 
\nonumber
\end{eqnarray}
In the limit $d \rightarrow 4$, we also 
confirm that the terms in curly bracket of 
right hand side results of (\ref{Fw2}, \ref{Ft2}) 
tend to zero. 
It means that the form factors 
always stay in finite in the limit.
\subsection{$H \rightarrow   
\gamma \gamma$ reduction}    
In order to reduce to $H \rightarrow   
\gamma \gamma$,  we take $M^2_Z \rightarrow 0$, 
and $\lambda^f_1 = e Q_f, 
\lambda^f_2, \lambda^f_3 \rightarrow 0$, 
the total amplitude of 
the decay $H \rightarrow Z \gamma$ 
is reduced to $H \rightarrow \gamma \gamma$. 
In detail, the results read
\begin{eqnarray}
\label{formfactors}
\mathcal{F}_{H\rightarrow \gamma\gamma}
(d; M_H^2, M_W^2, m_f^2) 
= \mathcal{F}^{(W)}_{H\rightarrow \gamma\gamma} 
(d; M_H^2, M_W^2)
+
\sum\limits_f N_C Q_f^2 
\mathcal{F}^{(f)}_{H\rightarrow \gamma\gamma}
(d; M_H^2, m_f^2).
\end{eqnarray}
Where the form factors are given 
\begin{eqnarray}
&&
\hspace{-0.6cm}
\dfrac{
\mathcal{F}^{(t)}_{H\rightarrow \gamma\gamma}
(d; M_H^2, m_t^2) 
}{\Gamma\left(2-d/2 \right) }
= \dfrac{(m_t^2)^{d/2 - 2} }{ (4 \pi)^{d/2} }
\Bigg\{ -\dfrac{8m_t^2}{M_H^2} 
\Fh32\Fz{1, 1, 2-d/2}
{3/2,1} {\dfrac{M_H^2}{4m^2_t }}  
\label{formtopA}
\\
&& \hspace{2cm}
+ \left(4 - d\right)
\Fh32\Fz{1, 1,3-d/2}
{3/2,2} {\dfrac{M_H^2}{4m^2_t }} 
+
\dfrac{8m_t^2}{M_H^2} 
\Fh32\Fz{1, 1, 2-d/2}
{3/2,2} {\dfrac{M_H^2}{4m^2_t }} 
\Bigg\},                                  
\nonumber
\end{eqnarray}
and
\begin{eqnarray}
&&
\dfrac{
\mathcal{F}^{(W)}_{H\rightarrow \gamma\gamma}
(d; M_H^2, M_W^2)
}
{\Gamma\left(2-d/2 \right)}
= \dfrac{  (M_W^2)^{d/2 - 2}  }
{ (4 \pi)^{d/2} }
\Bigg\{
\hspace{0cm}
 4 \left( 4 - d \right) \Fh32\Fz{1, 1, 3-d/2}
{ 3/2, 2} {\dfrac{M_H^2}{4M_W^2}}  + \\
&&
\hspace{0.6cm}
+ \Bigg[ 2  + 4 \dfrac{M_W^2}{M_H^2} 
( d - 1) \Bigg] \left( \Fh32\Fz{1, 1, 2-d/2}
{ 3/2, 2} {\dfrac{M_H^2}{4M_W^2}} 
- \Fh32\Fz{2, 1, 2-d/2}
{ 3/2, 2} {\dfrac{M_H^2}{4M_W^2}} 
\right)
\hspace{0.1cm}
\Bigg\}
\nonumber
\\
&&
\hspace{0.3cm}
= \dfrac{  (M_W^2)^{d/2 - 2}  }
{ (4 \pi)^{d/2} } \Bigg\{
\hspace{0.2cm}
\Big(2+ \frac{M_H^2}{M_W^2} \Big) 
\Fh32\Fz{2, 1, 2-d/2}
{ 3/2, 2} {\dfrac{M_H^2}{4M_W^2}} 
\label{WloopA}
\\
&&
\hspace{0.6cm}
-\Bigg[4 + \frac{M_H^2}{M_W^2} + 4(d-1)
\frac{M_W^2}{M_H^2} \Bigg]
\Fh32\Fz{1, 1,2-d/2}
{3/2,1} {\dfrac{M_H^2}{4M_W^2}} 
\nonumber\\
&&
\hspace{0.6cm}
+\Bigg[2 + 4(d-1)\frac{M_W^2}{M_H^2} \Bigg]
\Fh32\Fz{1, 1, 2-d/2}
{ 3/2,2} {\dfrac{M_H^2}{4M_W^2}} 
-4(d-4)\; 
\Fh32\Fz{1, 1, 3-d/2}
{3/2,2} {\dfrac{M_H^2}{4M_W^2}} 
\Bigg\},\nonumber
\end{eqnarray}
To arrive at the last line result, 
we have already used the transformation 
for hypergeometric functions $_3F_2$ 
in Eq. (\ref{transformation_3F2}).
These agree with 
the results in~\cite{khiemcadj}.
\subsection{Numerical results}
In numerical results,
we set  $M_H = 125$ GeV, 
$M_Z = 91.2$ GeV, $m_t = 173.5$ GeV 
and $M_W = 80.4$ GeV. Our results
are generated by using 
package {\tt NumEXP}~\cite{Huang:2012qz} 
for numerical $\epsilon$-expansions of
hypergeometric functions. We first confirm 
two representations for the form factors in 
(\ref{Fw1}, \ref{Ft1}) and (\ref{Fw2}, \ref{Ft2})
at general $d$. It means that we verify 
numerically the Ward identity at general $d$.
In Tables~\ref{testtop}, \ref{testW}, we show
numerical checks for the form factors at general 
$d$. Two representations for the form factors 
are perfect agreement up to last digit 
for $3.5 \leq d \leq 5.5$. 
\begin{table}[ht]
\begin{center}
\begin{tabular}{l@{\hspace{5em}}l}
 \hline \hline
$d$ & $\mathcal{F}_{H \rightarrow Z \gamma} ^{(t)}  
(d; M_H^2, M_Z^2,m_t^2)$ in Eq.~(\ref{Ft1}) \\ 
&  $\mathcal{F}_{H \rightarrow Z \gamma} ^{(t)}  
(d; M_H^2, M_Z^2,m_t^2)$ in Eq.~(\ref{Ft2}) \\ \hline \hline
$3.5$  & $-0.00117666222408164570889597705142$ \\
       &  $-0.00117666222408164570889597705142$ \\ \hline
$4.5$  & $-0.0756076123635421878866551078159$ \\
       & $-0.0756076123635421878866551078159$ \\ \hline
$5.0$  & $-0.754001360017782779626359989943$ \\
       & $-0.754001360017782779626359989943$ \\ \hline  
$5.5$  & $-10.6345811567309032438825219401$ \\
       & $-10.6345811567309032438825219401$ \\  
       \hline \hline
\end{tabular}
\caption{\label{testtop} Numerical confirmations
for two representations of the form factors involving
to top-loop diagrams at arbitrary $d$.}
\end{center}
\end{table}
\begin{table}[ht]
\begin{center}
\begin{tabular}{l@{\hspace{5em}}l}  \hline \hline
$d$ & $\mathcal{F}_{H \rightarrow Z \gamma} ^{(W)}  
(d; M_H^2, M_Z^2,M_W^2)$ in Eq.~(\ref{Fw1}) \\ 
&  $\mathcal{F}_{H \rightarrow Z \gamma} ^{(W)}  
(d; M_H^2, M_Z^2,M_W^2)$ in Eq.~(\ref{Fw2}) \\ \hline \hline
$3.5$  &  $-0.00924203129694608232780754562475$ \\
       &  $-0.00924203129694608232780754562475$ \\ 
\hline
$4.5$  & $-0.211488266331639234594811276488$ \\
       & $-0.211488266331639234594811276488$ \\ 
\hline
$5.0$  & $-1.26786296363083047430009124220$ \\
       & $-1.26786296363083047430009124220$ \\ 
\hline  
$5.5$  & $-10.8040444333273283701507434992$ \\
       & $-10.8040444333273283701507434992$ \\ 
\hline\hline
\end{tabular}
\caption{\label{testW} Numerical confirmations for 
two representations for the form factors involving to 
$W$-loop diagrams at arbitrary $d$.}
\end{center}
\end{table}

We next perform higher-order $\epsilon$-expansion
for the form factors in this work up to $\epsilon^5$.
We also compare our results with \cite{Hue:2017cph} 
($F^{SM} _{21,W}$)
at $\epsilon^0$-terms. Our numerical results are 
shown in Eqs.~(\ref{testHueW}, \ref{testHuetop}). 
We find a perfect agreement 
between two results at $\epsilon^0$-expansion. It is 
important to note that higher-power $\epsilon$-expansions
for the form factors in this paper are our first results.
\begin{eqnarray}
&&\hspace{5.1cm}
F^{SM} _{21,W} = 
-0.0418477713507083034768633206537 \; \epsilon^0 
\nonumber\\
&& \hspace{6.65cm} 
+
\;
\mathcal{O}(\epsilon);               
\\
&&
\mathcal{F}_{H \rightarrow Z \gamma} ^{(W)} 
(d = 4 - 2\epsilon; M_H^2, M_Z^2,M_W^2)
=
-0.0418477713507083034768633206537 \; \epsilon^0
\nonumber 
\\
&&\hspace{6.65cm}
+ 0.260913488721110921277821252790 \; \epsilon^1
\nonumber \\
&&\hspace{6.65cm}
-0.849415964842831522240990065525 \; \epsilon^2
 \nonumber \\
&&\hspace{6.65cm}
+ 1.93196240724203383916822579654 \; \epsilon^3
\nonumber \\
&&\hspace{6.65cm}
-3.46717780533875010127157401115 \; \epsilon^4
\nonumber \\
&&\hspace{6.65cm}
+ 5.25914558345954670519178485415 \; \epsilon^5
\nonumber\\
&&\hspace{6.65cm} +\mathcal{O}(\epsilon^6).      
\label{testHueW}
\end{eqnarray}
\begin{eqnarray}
&&\hspace{5.1cm}
F^{SM} _{21,t} = 
-0.00894937919735623466782637004746 \; \epsilon^0
\nonumber \\
&&\hspace{6.65cm}
+ \; 
\mathcal{O}(\epsilon);   
\\
&&
\mathcal{F}_{H \rightarrow Z \gamma} ^{(t)}  
(d= 4 - 2\epsilon; M_H^2, M_Z^2,m_t^2)
= 
-0.00894937919735623466782637004746 \; \epsilon^0
\nonumber 
\\
&&\hspace{6.4cm}
+ 0.0742785979879735824790115497100 \; \epsilon^1
\nonumber 
\\
&&\hspace{6.4cm}
-0.315615957203796781182876228270 \; \epsilon^2
\nonumber\\
&&\hspace{6.4cm}
+ 0.917527446546694361353843959657 \; \epsilon^3
\nonumber 
\\
&&\hspace{6.4cm}
-2.05845003852606360149227809637 \; \epsilon^4
\nonumber 
\\
&&\hspace{6.4cm}
+ 3.81281647820690166355588887060 \; \epsilon^5
\nonumber\\
&& \hspace{6.45cm} 
+
\; \mathcal{O}(\epsilon^6).
\label{testHuetop}
\end{eqnarray}
\section{Conclusions}   
In this paper, we have discussed the 
alternative approach for evaluating one-loop 
Feynman integrals. In this method, tensor one-loop 
integrals are reduced to scalar one-loop 
functions with the shifted space-time dimension. 
Scalar one-loop integrals are solved analytically
with the help of generalized hypergeometric 
series. We have applied this method for 
computing one-loop contributions to Higgs decay
to $Z\gamma$. For the first time, 
we have presented the form factors that are valid 
in general space-time dimension. The method 
can be extended to evaluate one-loop contributions
to Higgs decay to $Zf\bar{f}$, $f\bar{f}\gamma$, 
etc., within the SM and many BSMs. \\

\noindent
{\bf Acknowledgment:}~
This research is funded by Vietnam 
National Foundation for Science and 
Technology Development (NAFOSTED) 
under grant number $103.01$-$2019.346$.
\section*{Appendix $A$: Hypergeometric%
 sereis }                        
Series of hypergeometric functions
$_3F_2$~\cite{Slater} are defined: 
\begin{eqnarray}
 \Fh32\Fz{a_1,a_2,a_3}{b_1,b_2}{z}
 = \sum\limits_{m=0}^{\infty} 
 \dfrac{ (a_1)_m(a_2)_m (a_3)_m}
 { (b_1)_m(b_2)_m} \dfrac{z^m}{m!}.
\end{eqnarray}
The Mellin-Barnes representation for $_3F_2$ 
is 
\begin{eqnarray}
\Fh32\Fz{a_1,a_2,a_3}{b_1,b_2}{z} 
= \frac{\Gamma(a_1)\Gamma(a_2)\Gamma(a_3)}
{\Gamma(b_1)\Gamma(b_2)}
\dfrac{1}{2\pi i}
\int\limits_{-i\infty}^{i\infty} ds 
\Gamma(-s)\frac{\Gamma(s+a_1)
\Gamma(s+a_2)\Gamma(s+a_3)}
{\Gamma(s+b_1)\Gamma(s+b_2)}
\left(-z\right)^s, 
\nonumber\\
\end{eqnarray}
provided that $|\mathrm{Arg}(-z)|<\pi$. 
The integration contour is chosen 
in such a way that 
the poles of $\Gamma(- s)$ and 
$\Gamma(\cdots + s)$ are 
well-separated. Analytic continuation 
of $_3F_2$ functions:
\begin{eqnarray}
\label{analytic3F2}
&&\hspace{-1cm}\Fh32\Fz{a_1,a_2,a_3}{b_1,b_2}{z}=
\frac{\Gamma(b_1) \Gamma(b_2)}{\Gamma(a_1) 
\Gamma(a_2) \Gamma(a_3)} 
\times
\\
&&\hspace{0.5cm}\times
\Bigg\{\frac{
\Gamma(a_2-a_1)\Gamma(a_3-a_1) \Gamma(a_1)}
{\Gamma(b_1-a_1)\Gamma(b_2-a_1) (-z)^{a_1}}
 \Fh32\Fz{a_1,1-b_1+a_1,1-b_2+a_1}
 {1-a_2+a_1,1-a_3+a_1}{\frac{1}{z}}
\nonumber \\
&& \hspace{1cm}
+\frac{\Gamma(a_1-a_2)\Gamma(a_3-a_2)\Gamma(a_2)}
{\Gamma(b_1-a_2)\Gamma(b_2-a_2)~(-z)^{a_2}}
 \Fh32\Fz{a_2,1-b_1+a_2,1-b_2+a_2}
 {1-a_1+a_2,1-a_3+a_2}{\frac{1}{z}}
\nonumber \\
&&\hspace{1cm}
+\frac{
\Gamma(a_1-a_3) \Gamma(a_2-a_3)
\Gamma(a_3)}{\Gamma(b_1-a_3)
\Gamma(b_2-a_3)~(-z)^{a_3}}
\Fh32\Fz{a_3,1-b_1+a_3,1-b_2+a_3}
{1-a_1+a_3,1-a_2+a_3}{\frac{1}{z}} \Bigg\}. 
\label{F32_analytic}
\nonumber
\end{eqnarray}
In this work, a useful transformation 
for $\Fh32$ functions is mentioned:
\begin{eqnarray}
&&\hspace{-0.6cm}
\Fh32\Fz{a_1, a_2, a_3}
{b_1, b_2} {z} 
= \dfrac{b_1 - a_1}{b_1}
\Fh32\Fz{a_1, a_2, a_3}
{b_1 + 1, b_2} {z} 
+ \dfrac{a_1}{b_1}
\Fh32\Fz{a_1 + 1, a_2, a_3}
{b_1 + 1, b_2} {z} .
\label{transformation_3F2}
\end{eqnarray}

\section*{Appendix $B$: Calculating master 
integrals}                            
Tensor one-loop three-point Feynman integrals
with rank $M$ appearing in the process 
$H \rightarrow Z \gamma$ are given as follows:
\begin{eqnarray}
&&J_{3, \mu _1 \mu _2 \ldots \mu _M} 
( d; \{\nu_1 , \nu_2 , \nu_3 \} ) 
\nonumber\\
&& \equiv  J_{3, \mu _1 \mu _2 \ldots \mu _M} 
(d;\{ \nu _1, \nu _2, \nu _3 \}; p_2^2, M_H^2, M^2)
= \int \dfrac{{\rm d}^d k}{i \pi ^{d/2}} \; 
\dfrac{k_{\mu _1} k_{\mu _2} \ldots k_{\mu _M}}{
P_1 ^{\nu _1} P_2 ^{\nu _2} P_3 ^{\nu _3}}. 
\end{eqnarray}
Where the inverse Feynman propagators 
are 
\begin{eqnarray}
P_1 &=& \left( k + q_2 \right)^2 
- M ^2 + i \rho , \\
P_2 &=& \left( k + p \right)^2 - M ^2 + i \rho,\\
P_3 &=& k^2 - M ^2 + i \rho.
\end{eqnarray}
The related kinematic invariant are
$q_1 ^2 = M_Z^2, q_2 ^2 = 0,$ 
and $p^2 = (q_1 + q_2)^2 = M_H^2$. 
In this paper, $p_2^2 = M_Z^2, 0$ 
and internal masses $M^2 = m_f^2, M_W^2$.

After presenting tensor one-loop three-point
integrals to scalar functions, we next apply 
IBP for scalar one-loop functions with the general
propagator indexes. We then arrive 
at the following system of equations
\begin{small}
\begin{eqnarray}
\begin{cases}
( d - 2 \nu _1 - \nu _2 - \nu _3 ) \mathbf{1}  
- \nu _2 \mathbf{1}^{-} \mathbf{2}^{+}  
- \nu _3 \mathbf{1}^{-} \mathbf{3}^{+} 
=  \nu _1 ( 2 M^2 ) \mathbf{1}^{+} 
+  \nu _2 (2 M^2 - q_1 ^2) \mathbf{2}^{+} 
+  \nu _3 (2 M^2 - q_2 ^2) \mathbf{3}^{+}, \\
( d - \nu _1 - 2 \nu _2 - \nu _3 ) \mathbf{1}  
- \nu _1 \mathbf{1}^{+} \mathbf{2}^{-}  
- \nu _3 \mathbf{2}^{-} \mathbf{3}^{+} 
= \nu _1 (2 M^2 - q_1 ^2) \mathbf{1}^{+}  
+  \nu _2 (2 M^2 ) \mathbf{2}^{+} 
+  \nu _3 (2 M^2 - p ^2) \mathbf{3}^{+},
\\
( d - \nu _1 - \nu _2 - 2 \nu _3 ) \mathbf{1}  
- \nu _1 \mathbf{1}^{+} \mathbf{3}^{-}  
- \nu _2 \mathbf{2}^{+} \mathbf{3}^{-} 
= \nu _1 (2 M^2 - q_2 ^2) \mathbf{1}^{+} 
+  \nu _2 (2 M^2 - p^2) \mathbf{2}^{+} 
+  \nu _3 (2 M^2) \mathbf{3}^{+}.
\end{cases}
\end{eqnarray}
\end{small}
Here, the standard notation 
for increasing and lowering operators 
is used
\begin{eqnarray}
 \mathbf{j}^{\pm} 
J_3 (d;\{ \nu_j \}) 
= J_3 (d;\{ \nu _j \pm 1\})
\end{eqnarray}
for $j=1,2,3$.

In the following paragraphs, we consider
master integrals
$J_3 (d;\{ \nu _1, \nu _2, \nu _3 \})$  
by solving the above system of equations 
in several special cases. 
In conclusions, the master integrals 
shown at Section \ref{method} 
are presented in terms of 
hypergeometric functions $_3F_2$ 
in this paper. 
\subsubsection*{
\underline{\bf Case 1: 
$\nu _1 = \nu _2 = \nu _3 = 1$} }
\begin{eqnarray}
&& \hspace{-0.5cm}
J_3 (d;\{ 1, 2, 1 \}; p_2^2, M_H^2, M^2)
= \dfrac{ 2 }{  (p_2^2 - M_H^2)} 
\Big[ J_2 (d; \{ 2, 1 \}, M_H^2, M^2) 
- J_2 (d; \{ 2, 1 \}, p_2^2, M^2) \Big],
\nonumber \\
\end{eqnarray}
\begin{eqnarray}
&&
\hspace{-0.3cm}
J_3 (d;\{ 2, 1, 1 \}; p_2^2, M_H^2, M^2)= \\
&& 
= \dfrac{ (d-4) M_H^2 }{2 M^2 (M_H^2 - p_2^2)} 
J_3 (d;\{ 1, 1, 1 \}; p_2^2, M_H^2, M^2)  
+ \dfrac{ 2}{ (p_2^2 - M_H^2)} 
J_2 (d; \{ 2, 1 \}, 0, M^2) 
\nonumber \\
&&
+ \dfrac{ M_H^2 (4 M^2 -  M_H^2)}
{ M^2 (M_H^2-p_2^2)^2} 
J_2 (d; \{ 2, 1 \}, M_H^2, M^2)  
+ \dfrac{ p_2^2 M_H^2  - 2 M^2 
(p_2^2  + M_H^2 ) }{ 
M^2 (M_H^2-p_2^2)^2} 
J_2 (d; \{ 2, 1 \}, p_2^2, M^2),
\nonumber
\end{eqnarray}
\begin{eqnarray}
&& 
\hspace{-0.3cm}
J_3 (d;\{ 1, 1, 2 \}; p_2^2, M_H^2, M^2) = 
\\
&&
= \dfrac{ (d - 4) p_2^2 }{2 M^2 (p_2^2 - M_H^2)} 
J_3 (d;\{ 1, 1, 1 \}; p_2^2, M_H^2, M^2) 
+ \dfrac{ 2 }{ (M_H^2 - p_2^2)} 
J_2 (d; \{ 2, 1 \}, 0, M^2) 
\nonumber\\
&&
+ \dfrac{ p_2^2 M_H^2  - 2 M^2 (M_H^2 + p_2^2 ) }{ 
M^2 (M_H^2-p_2^2)^2} 
J_2 (d; \{ 2, 1 \}, M_H^2, M^2) 
+ \dfrac{ p_2^2 (4 M^2 - p_2^2)}{ 
M^2 (M_H^2-p_2^2)^2} 
J_2 (d; \{ 2, 1 \}, p_2^2, M^2). \nonumber
\end{eqnarray}
\subsubsection*{\underline{\bf Case 2: 
$\nu _1 = 1 , \nu _2 = 2 , \nu _3 = 1$ }}
\begin{eqnarray}
&&\hspace{-2.4cm}
J_3 (d;\{ 1, 3, 1 \}; p_2^2, M_H^2, M^2) 
= \\
&=&
\dfrac{1}{2 (p_2^2-M_H^2)} 
\Big[ J_2 (d; \{ 2, 2 \}, M_H^2, M^2) 
- J_2 (d; \{ 2, 2 \}, p_2^2, M^2) \Big]
\nonumber\\
&&
+ \dfrac{1}{ (p_2^2-M_H^2)}  
\Big[ J_2 (d; \{ 3, 1 \}, M_H^2, M^2) 
- J_2 (d; \{ 3, 1 \}, p_2^2, M^2) \Big],
\nonumber
\end{eqnarray}
\begin{eqnarray}
&& \hspace{-1cm}
J_3 (d;\{ 2, 2, 1 \}; p_2^2, M_H^2, M^2) 
= \\
&=&  \dfrac{ (4 - d) }{
2 M^2 (M_H^2-p_2^2)} 
J_3 (d;\{ 1, 1, 1 \}; p_2^2, M_H^2, M^2)
\nonumber
\\
&&
+ \dfrac{ M_H^2 \left(4 M^2  - M_H^2 \right)}{
2 M^2 (M_H^2-p_2^2)^2} 
\Big[ J_2 (d; \{ 2, 2 \}, M_H^2, M^2) 
+ 2 J_2 (d; \{ 3, 1 \}, M_H^2, M^2) 
\Big]
\nonumber \\
&&
+ \dfrac{ p_2^2 M_H^2  -2 M^2 (M_H^2 + p_2^2)}{
2 M^2 (M_H^2-p_2^2)^2} 
\Big[ J_2 (d; \{ 2, 2 \}, p_2^2, M^2) 
+ 2 J_2 (d; \{ 3, 1 \}, p_2^2, M^2) \Big]
\nonumber\\
&&
+ \dfrac{  (6 - d) M_H^2}{ M^2 (M_H^2-p_2^2)^2} 
J_2 (d; \{ 2, 1 \}, M_H^2, M^2) 
+ \dfrac{ (d - 5) M_H^2 - p_2^2}{ 
M^2 (M_H^2-p_2^2)^2} 
J_2 (d; \{ 2, 1 \}, p_2^2, M^2)
, \nonumber
\end{eqnarray}
\begin{eqnarray}
&&\hspace{-1.0cm}
J_3 (d;\{ 1, 2, 2 \}; p_2^2, M_H^2, M^2) 
=
\\
&&\hspace{-0.8cm}
=
\hspace{0.1cm}
\dfrac{ (d-4)}{2 M^2 (M_H^2-p_2^2)} 
J_3 (d;\{ 1, 1, 1 \}; p_2^2, M_H^2, M^2) 
\nonumber
\\
&&
+ \dfrac{ p_2^2 M_H^2 - 2 M^2 ( M_H^2 + p_2^2)}{
2 M^2 (M_H^2-p_2^2)^2} 
\Big[ J_2 (d; \{ 2, 2 \}, M_H^2, M^2) 
+ 2 J_2 (d; \{ 3, 1 \}, M_H^2, M^2) 
\Big]
\nonumber\\
&&
+ \dfrac{ p_2^2 
\left(4 M^2 - p_2^2\right)}{
2 M^2 (M_H^2-p_2^2)^2} 
\Big[ J_2 (d; \{ 2, 2 \}, p_2^2, M^2) 
+ 2 J_2 (d; \{ 3, 1 \}, p_2^2, M^2) 
\Big]
\nonumber
\\
&&
+ \dfrac{ (d-5) p_2^2 - M_H^2}{ 
M^2 (M_H^2-p_2^2)^2} 
J_2 (d; \{ 2, 1 \}, M_H^2, M^2) 
+ \dfrac{  (6 - d) p_2^2}{ M^2 (M_H^2-p_2^2)^2} 
J_2 (d; \{ 2, 1 \}, p_2^2, M^2)
.
\nonumber
\end{eqnarray}
\section*{Appendix $C$: One-loop amplitudes%
~for $H\rightarrow Z\gamma$}           
We present detailed calculations for the 
decay amplitude of $H\rightarrow Z\gamma$ 
in unitary gauge in this appendix. 
All couplings involving the decay process 
are listed in Table~\ref{couplings table}. 
\begin{table}
\begin{center}
\begin{tabular}{ll} 
\hline\hline
\textbf{Vertices} & \textbf{Couplings}   \\ \hline\hline
$H W_\mu W_\nu $ & $i g M_W g^{\mu \nu}$  \\ 
$Z_\mu (k_1) W_\nu (k_2) W_\lambda (k_3)$ 
& $- i g \cos \theta_W 
\left[(k_1 - k_2)_\lambda g^{\mu \nu} 
+ (k_2 - k_3)_\mu g^{\nu \lambda} 
+ (k_3 - k_1)_\nu g^{\lambda \mu}  \right]$ \\ 
$A_\mu (k_1) W_\nu (k_2) W_\lambda (k_3)$ 
& $- i e \left[(k_1 - k_2)_\lambda g^{\mu \nu} 
+ (k_2 - k_3)_\mu g^{\nu \lambda} 
+ (k_3 - k_1)_\nu g^{\lambda \mu}  \right]$ \\ 
$A_\mu Z_\nu W_\alpha W_\beta$ 
& $- i e g \cos \theta_W 
\left[ 2 g^{\mu \nu} g^{\alpha \beta} 
- g^{\mu \alpha} g^{\nu \beta} 
- g^{\mu \beta} g^{\nu \alpha} \right]$  
\\ 
$H Z_\mu Z_\nu $ 
& $i \, g M_Z g^{\mu \nu} / \cos \theta_W$ 
\\ 
$H t t$ 
& $ - i \, g m_t / 2 M_W$  \\ 
$t t Z_\mu $ & $
i \, \left( \lambda^t_1 + \lambda^t_2 \right)\gamma ^\mu  
+ i \, \lambda^t_3 \gamma ^\mu \gamma ^5$ 
\\ 
$t t A_\mu $ & $i e Q_t \gamma ^\mu$  
\\ \hline\hline
\end{tabular}
\end{center}
\caption{Couplings involving 
the decay $H \rightarrow Z \gamma$. 
In our notation, $\lambda^t_1 =  e Q_t, 
\;\lambda^t_2 = \dfrac{g}{2 \cos \theta _W} 
\left( I^3 _t - 2 Q_t \sin ^2 \theta _W 
- \, 2 Q_t \sin \theta_W \cos \theta_W \right)$, 
and $\lambda^t_3 = 
-  \dfrac{g}{2 \cos \theta _W} I^3 _t$.
Where $I^{3}_t$ and $Q_t$ are iso-spin 
and electric charge of 
top quarks in the loops. 
The term $i \, \lambda^t_1 \gamma ^\mu$ 
should be the coupling of photon to top quarks.
\label{couplings table}}
\end{table}
The decay amplitude $H \rightarrow Z \gamma$ 
of top-loop diagrams is expressed
as follows:
\begin{eqnarray}
&&\hspace{-0.5cm}
i \mathcal{A}^{(T)}_{H \rightarrow Z \gamma}
=
\hspace{0.1cm}
- \dfrac{e Q_t g m_t^2}{(4 \pi)^{d/2} M_W} 
\left( \lambda^t_1 
+ \lambda^t_2 \right)
\int \frac{{\rm d}^d k}{i \pi ^{d/2}} \, 
\dfrac{\varepsilon ^{\mu *} _1 (q_1) 
\varepsilon ^{\nu *} _2 (q_2) }{
P_1 P_2 P_3} 
\times
\\
&&\hspace{-0.5cm}
\times
\Bigg\{
16 k^{\mu } k^{\nu } 
+ 8 k^{\nu } q_1^{\mu } 
+ 16 k^{\nu } q_2^{\mu } 
+ 4 q_1^{\nu } q_2^{\mu } 
- g^{\mu \nu} \Big[ 
8 (k \cdot q_2) 
+ 4 k^2 
+  \left(2 M_H^2 - 2 M_Z^2 
- 4 m_t^2 \right)
\Big] \Bigg\}. 
\nonumber
\end{eqnarray}
Where the coefficient factors 
are written in terms of the master
integrals:
\begin{eqnarray}
&&\hspace{-0.8cm}
\int  \dfrac{{\rm d}^d k}{
i \pi^{d/2}}\, \dfrac{k^{\mu } k^{\nu }}{
P_1 P_2 P_3} 
=
\hspace{0.1cm}
\dfrac{- g^{\mu \nu}}{2} 
J_3 (d+2;\{1,1,1\};M_Z^2,M_H^2,m_t^2) 
 \\
&&\hspace{2.6cm}
+ \; q_1^\mu q_1^\nu 
J_3 (d+4;\{1,3,1\};M_Z^2,M_H^2,m_t^2)
\nonumber \\
&&
\hspace{2.6cm}
+ \; q_2^\mu q_1^\nu \Big[ 
J_3 (d+4;\{2,2,1\};M_Z^2,M_H^2,m_t^2) 
+ J_3 (d+4;\{1,3,1\};M_Z^2,M_H^2,m_t^2) \Big],
\nonumber\\
&&\hspace{-0.8cm}
\int  \dfrac{{\rm d}^d k}{i \pi^{d/2}}\, 
\dfrac{ k^{\nu } q_1^{\mu } }{
P_1 P_2 P_3}
=
\hspace{0.1cm}
q_1^{\mu } q_1^{\nu} 
J_3 (d+2;\{1,2,1\};M_Z^2,M_H^2,m_t^2)
,
\\
&&\hspace{-0.8cm}
\int  \dfrac{{\rm d}^d k}{i \pi^{d/2}}\, 
\dfrac{ k^{\nu } q_2^{\mu } }{
P_1 P_2 P_3} 
=
\hspace{0.1cm}
q_2^{\mu } q_1^{\nu} 
J_3 (d+2;\{1,2,1\};M_Z^2,M_H^2,m_t^2)
,
\\
&&\hspace{-0.8cm}
\int  \dfrac{{\rm d}^d k}{i \pi^{d/2}}\, 
\dfrac{ q_1^{\nu } q_2^{\mu } }{
P_1 P_2 P_3} 
=
\hspace{0.1cm}
q_2^{\mu } q_1^{\nu } 
J_3 (d;\{1,1,1\};M_Z^2,M_H^2,m_t^2)
,
\\
&&\hspace{-0.8cm}
\int  \dfrac{{\rm d}^d k}{i \pi^{d/2}}\, 
\dfrac{ k \cdot q_2 }{
P_1 P_2 P_3} 
=
\hspace{0.1cm}
\frac{1}{2} \Big[ 
J_2 (d;\{ 1,1 \}; M_H^2, m_t^2) 
- J_2 (d;\{ 1,1 \}; M_Z^2, m_t^2) \Big]
,
\\
&&\hspace{-0.8cm}
\int  \dfrac{{\rm d}^d k}{i \pi^{d/2}}\, 
\dfrac{ k ^2 }{P_1 
P_2 P_3} 
=
\hspace{0.1cm}
J_2 (d;\{ 1,1 \}; M_Z^2, m_t^2) 
+ m_t^2 J_3 (d;\{ 1,1,1 \}; 
M_Z^2, M_H^2, m_t^2) .
\end{eqnarray}
For the $W$ boson loops contributions, 
the decay amplitude is written 
\begin{eqnarray}
&&\hspace{-0.5cm}
i \mathcal{A}^{(W)}_{H \rightarrow Z \gamma}
=
\hspace{0.1cm}
\dfrac{i e g^2 \cos \theta_W }{ 
(4 \pi)^{d/2} M_W^5 } 
\int \dfrac{{\rm d}^d k}{
i \pi ^{d/2}} \; 
\varepsilon ^{\mu *} _1 (q_1) 
\varepsilon ^{\nu *} _2 (q_2) 
\times
\\
&&\hspace{1.2cm}
\times
\Big[ i \mathcal{A}_1 g^{\mu \nu } 
+ i \mathcal{A}_2 k^\mu k^\nu 
+ i \mathcal{A}_3 k^\mu q_1^\nu 
+ i \mathcal{A}_4 q_2^\mu k^\nu 
+ i \mathcal{A}_5 q_2^\mu q_1^\nu 
+ i \mathcal{A}_6 q_1^\mu k^\nu 
+ i \mathcal{A}_7 q_1^\mu q_1^\nu 
\Big].
\nonumber
\end{eqnarray}
Where the coefficient factors 
are presented in terms of the master 
integrals in detail:
\begin{eqnarray}
&&\hspace{-0.8cm}
\int \dfrac{{\rm d}^d k}{
i \pi ^{d/2}} \;  
(i \mathcal{A}_1 g^{\mu \nu} )
= \hspace{0.1cm}
\int \dfrac{{\rm d}^d k}{
i \pi ^{d/2}} \;
\Bigg\{  \hspace{0.1cm}
\frac{1}{P_1 P_2 P_3} 
\Big[ 2 M_W^4 
\Big( M_Z^2 - 4 M_W^2 \Big) 
\Big( M_H^2 - M_Z^2 \Big) \Big]
\\
&&\hspace{1.5cm}
+ \frac{1}{P_2 
P_3} 
\Big[ 
2 M_W^4 \Big( M_Z^2 - M_H^2 \Big)
 + M_H^2 M_W^2 M_Z^2
- \frac{M_H^4 M_W^2}{2} 
\Big]
\nonumber
\\
&&\hspace{1.5cm}
+ \frac{1}{P_2 P_3} 
\Big[ 
-P_1 
\Big( M_H^2 M_W^2 + 2 M_W^4 \Big) 
+ 2 M_W^6 \Big( 1 - d \Big)
\Big]
+ \frac{1}{P_1} 
\Big(- 2 M_W^4 \Big)
\nonumber
\\
&&\hspace{1.5cm}
+ \Big( \frac{1}{P_2} + \frac{1}{P_3} \Big)
\Big[ \frac{M_W^2}{2}  \Big( 2 P_1
- P_2 - P_3 \Big)
+ M_W^2 \Big( 2 M_W^2 + M_H^2 - M_Z^2 \Big) 
\Big] 
\hspace{0.2cm}   \Bigg\}
\; g^{\mu \nu}
\nonumber \\
&&\hspace{0cm}
=
\hspace{0.1cm}
\Big\{ 
\hspace{0.1cm}
2 M_W^4 \Big( M_Z^2 - 4 M_W^2 \Big) 
\Big( M_H^2 - M_Z^2 \Big)
J_3 (d;\{ 1, 1, 1 \}; 
M_Z^2, M_H^2, M_W ^2)
\\
&&\hspace{0.5cm}
+ \Big[ 2 M_W^6 \Big( 1 - d \Big) 
+M_H^2 M_W^2 M_Z^2 
+ M_W^4 \Big( 2 M_Z^2 - M_H^2 \Big)
\Big] 
J_2 ( d;\{ 1, 1 \}; M_H^2, M_W^2 )
\nonumber
\\
&&\hspace{0.5cm}
+ M_W^2 M_Z^2 \Big( M_H^2 + 2 M_W^2 \Big) 
J_2 ( d + 2;\{ 2, 1 \}; M_H^2, M_W^2 ) 
-M_W^2 M_Z^2\; J_1 (d;\{ 1 \}; M_W^2) 
\hspace{0.1cm}
\Big\} \; g^{\mu \nu}, 
\nonumber
\end{eqnarray}
\begin{eqnarray}
&&\hspace{-0.8cm}
\int \dfrac{{\rm d}^d k}{i \pi ^{d/2}} \; 
( i \mathcal{A}_2 k^\mu k^\nu )
=
\\
&&\hspace{-1.0cm}
=
\hspace{0.1cm}
\int \dfrac{{\rm d}^d k}{i \pi ^{d/2}} \;
\Big\{ 
\frac{k^\mu k^\nu}{P_1 
P_2 P_3} 
\Big[ 4 M_W^4 \Big( M_H^2 - M_Z^2 \Big)
-2 M_H^2 M_W^2 M_Z^2
+ 8 M_W^6 \Big( d - 1 \Big)
 \Big] 
+\frac{k^\mu k^\nu}{
P_1 P_3} 
\Big( 2 M_W^2 M_Z^2 \Big)
\Big\}
\nonumber
\\
&&\hspace{-1.0cm}
=\hspace{0.1cm}\nonumber
\Big\{\hspace{0.1cm}
\Big[ 2 M_W^4 \Big( M_Z^2 - M_H^2 \Big)
+  M_H^2 M_W^2 M_Z^2 + 4 M_W^6 \Big(1-d\Big) \Big] 
J_3 (d + 2;\{ 1,1,1 \}; 
M_Z^2, M_H^2, M_W^2) \nonumber \\
&&\hspace{-0.2cm}
- M_W^2 M_Z^2\;
J_2 (d + 2;\{ 1, 1 \}; 0, M_W^2)
\hspace{0.1cm} \Big\} \; g ^{\mu \nu}
\\
&&\hspace{-0.8cm}
+\Big\{\hspace{0.1cm}
\Big[ 4 M_W^4 \Big( M_H^2 - M_Z^2 \Big)
-2 M_H^2 M_W^2 M_Z^2
+ 8 M_W^6 \Big( d - 1 \Big)
 \Big] \times\nonumber \\
&&\times \Big[ J_3 (d + 4;\{ 2,2,1 \}; 
M_Z^2, M_H^2, M_W^2) 
+ J_3 (d + 4;\{ 1,3,1 \}; 
M_Z^2, M_H^2, M_W^2) \Big]
\hspace{0.1cm}
\Big\}\; q_2^\mu q_1^\nu
\nonumber\\
&&\hspace{-0.8cm}
+
\Big\{
\hspace{0.1cm}
\Big[ 4 M_W^4 \Big( M_H^2 - M_Z^2 \Big)
-2 M_H^2 M_W^2 M_Z^2
+ 8 M_W^6 \Big( d - 1 \Big)  \Big] 
J_3 (d + 4;\{ 1,3,1 \}; 
M_Z^2, M_H^2, M_W^2)
\hspace{0.1cm}
\Big\}q_1^\mu q_1^\nu, \nonumber
\end{eqnarray}
\begin{eqnarray}
&&\hspace{-0.8cm}
\int \dfrac{{\rm d}^d k}{i \pi ^{d/2}} \; 
(i \mathcal{A}_3 k^\mu q_1^\nu )= \\
&=&
\int \dfrac{{\rm d}^d k}{i \pi ^{d/2}} \;
\Big\{ \frac{k^\mu q_1^\nu}{P_1 P_2} 
(4 M_W^4-2 M_W^2 M_Z^2) 
- \frac{k^\mu q_1^\nu}{P_2 P_3} 
\left(\frac{M_H^2 M_W^2}{2}+7 M_W^4 \right) 
+ \frac{M_W^2}{2}
\left(\frac{k^\mu q_1^\nu}{P_2} 
+\frac{k^\mu q_1^\nu}{P_3} \right)
\Big\} \nonumber\\
&=& \Big\{
\Big(2 M_W^2 M_Z^2 - 4 M_W^4 \Big) 
J_2 (d;\{ 1, 1 \}; M_Z^2, M_W^2)
\nonumber\\
&& - \left( \frac{M_H^2 M_W^2}{2} 
+ 7 M_W^4 \right) 
J_2 (d + 2;\{ 2, 1 \}; M_H^2, M_W^2)  
-  \frac{M_W^2}{2} J_1 (d; \{ 1 \}; M_W^2)
\Big\} \; q_2^\mu q_1^\nu  
\nonumber\\
&& +
\Big\{ \Big( 4 M_W^4 - 2 M_W^2 M_Z^2 \Big) 
J_2 (d+2;\{ 2, 1 \}; M_Z^2, M_W^2)
\nonumber \\
&& - \left( \frac{M_H^2 M_W^2}{2} + 7 M_W^4 \right) 
J_2 (d + 2;\{ 2, 1 \}; M_H^2, M_W^2)  
- \left( \frac{M_W^2}{2} \right) 
J_1 (d; \{ 1 \}; M_W^2)
\Big\} \; q_1^\mu q_1^\nu,
\nonumber 
\end{eqnarray}
\begin{eqnarray}
&&\hspace{-0.8cm}
\int \dfrac{{\rm d}^d k}{i \pi ^{d/2}} \; 
( i \mathcal{A}_4 q_2^\mu k^\nu ) =         \\
&=& \int\dfrac{{\rm d}^d k}{i \pi ^{d/2}} \;
\Big\{ \frac{q_2^\mu k^\nu}{P_1 P_2 P_3} 
\Big[ 4 M_W^4 \Big( M_H^2 - M_Z^2 \Big)
-2 M_H^2 M_W^2 M_Z^2
+ 8 M_W^6 \Big( d - 1 \Big)  \Big] 
\nonumber\\
&& +\frac{q_2^\mu k^\nu}{P_1 P_3} 
\Big( 2 M_W^2 M_Z^2-4 M_W^4 \Big) 
+\frac{q_2^\mu k^\nu}{P_2 P_3} 
(\frac{M_H^2 M_W^2}{2} +7 M_W^4)
-\frac{M_W^2}{2}
\Big( \frac{q_2^\mu k^\nu}{P_3} 
+
\frac{q_2^\mu k^\nu}{P_2} \Big)
\Big\}
\nonumber \\
&=&\Big\{
\Big[ 4 M_W^4( M_H^2 - M_Z^2)
-2 M_H^2 M_W^2 M_Z^2
+ 8 M_W^6(d-1)
 \Big]
J_3 (d + 2;\{ 1,2,1 \}; 
M_Z^2, M_H^2, M_W^2)
\nonumber \\
&&
+ \left( \frac{M_H^2 M_W^2}{2}
+7 M_W^4 \right) 
J_2 (d + 2;\{ 2,1 \}; M_H^2, M_W^2) 
+\frac{M_W^2}{2} \; J_1 (d;\{ 1 \}; M_W^2)
\Big\}\; q_2^\mu q_1^\nu, 
\nonumber
\end{eqnarray}
\begin{eqnarray}
&& 
\hspace{-2cm}
\int \dfrac{{\rm d}^d k}{i \pi ^{d/2}} \; 
( i \mathcal{A}_5 q_2^\mu q_1^\nu )     
=
\int \dfrac{{\rm d}^d k}{i \pi ^{d/2}} \;
\Big\{ \frac{( 16 M_W^6-4 M_W^4 M_Z^2)}
{P_1  P_2 P_3}  
+\frac{( 4 M_W^4-2 M_W^2 M_Z^2)}{P_1 P_2} 
\Big\}\; q_2^\mu q_1^\nu
\nonumber \\
&&\hspace{1.5cm}
=\Big\{ ( 16 M_W^6-4 M_W^4 M_Z^2) 
J_3 (d;\{ 1,1,1 \}; M_Z^2, M_H^2, M_W^2)
\nonumber\\
&& \hspace{3cm}
+ ( 4 M_W^4-2 M_W^2 M_Z^2 ) 
J_2 (d;\{ 1,1 \}; M_Z^2, M_W^2)
\Big\} \; q_2^\mu q_1^\nu,
\end{eqnarray}
\begin{eqnarray}
&&\hspace{-0.8cm}
\int \dfrac{{\rm d}^d k}{i \pi ^{d/2}} \; 
( i \mathcal{A}_6 q_1^\mu k^\nu ) =
\\ 
&=& \int \dfrac{{\rm d}^d k}{i \pi ^{d/2}} \;
\Big\{ 
\dfrac{q_1^\mu k^\nu}{P_1 P_2 P_3} 
[2 M_W^4 ( M_H^2 - M_Z^2 ) 
-M_H^2 M_W^2 M_Z^2 + 4 M_W^6 (d - 1)]
+ \frac{q_1^\mu k^\nu}{P_2} 
\left( \frac{M_W^2}{2} \right)
\nonumber
\\
&&+ \frac{q_1^\mu k^\nu}{P_2 P_3} 
\left( 3 M_W^4-\frac{3 M_H^2 M_W^2}{2} \right) 
+ \frac{q_1^\mu k^\nu}{P_1 P_3} 
\Big[ - M_W^2 P_2 
+ M_W^2 \Big( M_H^2 + M_Z^2 \Big)
\Big]
+ \frac{q_1^\mu k^\nu}{ P_3} 
\left( \dfrac{3 M_W^2}{2} \right)
\Big\}
\nonumber \\
&=& \Big\{
\Big[ 2 M_W^4 ( M_H^2 - M_Z^2)
-M_H^2 M_W^2 M_Z^2
+ 4 M_W^6( d - 1) \Big]
J_3 (d + 2;\{ 1,2,1 \}; 
M_Z^2, M_H^2, M_W^2)
\nonumber \\
&& + \left( 3 M_W^4
-\frac{3 M_H^2 M_W^2}{2} \right) 
J_2 (d + 2;\{  2,1 \}; M_H^2, M_W^2)  
\nonumber \\
&& + 
\left(-\frac{M_W^2}{2} \right) 
\Big[ J_1 (d;\{ 1 \}; M_W^2) 
- 2 J_2 (d + 2;\{ 1,1 \}; 0, M_W^2)
\Big] \Big\} \; q_1^\mu q_1^\nu, 
\nonumber
\end{eqnarray}
\begin{eqnarray}
&&\hspace{-0.8cm}
\int \dfrac{{\rm d}^d k}{i \pi ^{d/2}} \; 
( i \mathcal{A}_7 q_1^\mu q_1^\nu )
= \\
&=& \int \dfrac{{\rm d}^d k}{i \pi ^{d/2}} \;
\Big\{ \frac{1}{P_2 P_3} (-M_H^2 M_W^2-2 M_W^4)
+ \dfrac{1}{P_1 P_2} (2 M_W^4-M_W^2 M_Z^2) 
+  M_W^2 \Big(\dfrac{1}{P_1} +\dfrac{1}{P_3} \Big) 
\Big\}\; q_1^\mu q_1^\nu
\nonumber\\
&&
\hspace{-0.7cm}
= \Big\{( -M_H^2 M_W^2-2 M_W^4 ) 
J_2 (d;\{ 1,1 \}; M_H^2, M_W^2) \nonumber\\
&& + (2 M_W^4-M_W^2 M_Z^2) 
J_2 (d;\{ 1,1 \}; M_Z^2, M_W^2)  
+ 2 M_W^2 J_1 (d;\{ 1 \}; M_W^2) 
\Big\}\; q_1^\mu q_1^\nu
. \nonumber
\end{eqnarray}
\section*{Appendix $D$: Form factors
~at general $d$ \label{form_factors}}
The form factors in Eq. (\ref{total amplitude})
are
\begin{eqnarray}
&&\hspace{-0.6cm}
F^{(t)}_{00} 
= - \dfrac{e Q_t g m_t^2}{(4 \pi)^{d/2} M_W} 
\left( \lambda^{(t)}_1 + \lambda^{(t)}_2 \right) 
\Big[ \left(2  M_Z^2 - 2  M_H^2 \right) 
J_3 (d;\{ 1,1,1 \}; M_Z^2, M_H^2, m_t^2)
\\
&&
\hspace{4cm} 
- 8 J_3 (d+2;\{1,1,1\};M_Z^2,M_H^2,m_t^2) 
- 4 J_2 (d;\{ 1,1 \}; M_H^2, m_t^2) 
\Big] , \nonumber 
\\
&&\hspace{-0.6cm}
F^{(t)}_{11} 
= - \dfrac{e Q_t g m_t^2}{(4 \pi)^{d/2} M_W} 
\left( \lambda^{(t)}_1 + \lambda^{(t)}_2 \right) 
\Big[ 
\hspace{0.2cm}
16 J_3 (d+4;\{ 1,3,1 \};M_Z^2,M_H^2,m_t^2)
\\
&&
\hspace{5.6cm} 
+ 8 J_3 (d+2;\{ 1,2,1 \};M_Z^2,M_H^2,m_t^2)
\hspace{0.2cm}
\Big] , \nonumber 
\\
&&\hspace{-0.6cm}
F^{(t)}_{21} 
= - \dfrac{e Q_t g m_t^2}{(4 \pi)^{d/2} M_W} 
\left( \lambda^{(t)}_1 + \lambda^{(t)}_2 \right) 
\Big[ 16 J_3 (d+4;\{2,2,1\};M_Z^2,M_H^2,m_t^2)
\\
&&
\hspace{2.5cm}
+ 8 J_3 (d+2;\{1,2,1\};M_Z^2,M_H^2,m_t^2)
+ 4 J_3 (d;\{1,1,1\};M_Z^2,M_H^2,m_t^2)
\Big] , \nonumber
\\
&&\hspace{-0.6cm}
F^{(t)}_5 = 0,
\end{eqnarray}
and 
\begin{eqnarray}
&&\hspace{-0.6cm}
F^{(W)}_{00} 
=  \dfrac{ e g^2 \cos \theta_W }
{ (4 \pi)^{d/2} M_W^5 } 
\Bigg\{ \hspace{0.1cm}
2 M_W^4 (  M_Z^2 - 4 M_W^2 ) 
( M_H^2 - M_Z^2 ) 
J_3 (d;\{ 1, 1, 1 \}; M_Z^2, M_H^2, M_W^2) 
\\
&&
\hspace{1.0cm}
+ \Big[ 2 M_W^4 ( M_Z^2 - M_H^2 ) 
+  M_H^2 M_W^2 M_Z^2 
- 4 M_W^6 (d - 1) \Big] 
J_3 (d + 2;\{ 1,1,1 \}; M_Z^2, M_H^2, M_W^2)
\nonumber
\\
&&
\hspace{1.0cm}
+ \Big[ 2 M_W^4 ( M_H^2 - M_Z^2 ) 
-  M_H^2 M_W^2 M_Z^2 
+ 4 M_W^6 (d - 1) \Big] 
J_2 ( d + 2;\{ 2, 1 \}; M_H^2, M_W^2 )
\hspace{0.3cm}
\Bigg\} , \nonumber
\end{eqnarray}
\begin{eqnarray}
&&\hspace{-0.6cm}
F^{(W)}_{11} 
=  \dfrac{ e g^2 \cos \theta_W }
{ (4 \pi)^{d/2} M_W^5 } 
\times
\\
&&
\times 
\Big\{
\hspace{0.1cm}
\Big[ 4 M_W^4 ( M_H^2 - M_Z^2 ) 
- 2 M_H^2 M_W^2 M_Z^2 
+ 8 M_W^6 (d - 1) \Big] 
J_3 (d + 4;\{ 1,3,1 \}; M_Z^2, M_H^2, M_W^2)
\nonumber
\\
&&
\hspace{0.5cm}
+ \Big[ 2 M_W^4 ( M_H^2 - M_Z^2 ) 
-  M_H^2 M_W^2 M_Z^2 
+ 4 M_W^6 (d - 1) \Big] 
J_3 (d + 2;\{ 1,2,1 \}; M_Z^2, M_H^2, M_W^2)
\hspace{0.1cm}
\Big\} , \nonumber
\end{eqnarray}
\begin{eqnarray}
&&\hspace{-0.6cm}
F^{(W)}_{21} 
=  \dfrac{ e g^2 \cos \theta_W }
{ (4 \pi)^{d/2} M_W^5 } 
\Big\{
\hspace{0.1cm}
4 M_W^4 
\left( 4 M_W^2 - M_Z^2 \right) 
J_3 (d;\{ 1,1,1 \}; M_Z^2, M_H^2, M_W^2)
\\
&&
\hspace{0.5cm}
+ \Big[2 M_W^4 (M_H^2 - M_Z^2) 
-  M_H^2 M_W^2 M_Z^2 
+ 4 M_W^6 \, (d - 1) \Big] 
J_3 (d + 2;\{ 1,2,1 \}; M_Z^2, M_H^2, M_W^2)
\nonumber
\\
&&
\hspace{0.5cm}
+ \Big[4 M_W^4 (M_H^2 - M_Z^2) - 2 M_H^2 M_W^2 M_Z^2 
+ 8 M_W^6 \, (d - 1) \Big]
J_3 (d + 4;\{ 2,2,1 \}; M_Z^2, M_H^2, M_W^2)
\hspace{0.1cm}
\Big\} . \nonumber
\end{eqnarray}

\end{document}